# Performance of using Mel-Frequency Cepstrum Based Features in Nonlinear Classifiers for Phonocardiography Recordings

Ibrahim Ozkan
*Hacettepe University*
*Electrical and Electronics Engineering*
Ankara, Turkiye
ibrahimozkan@hacettepe.edu.tr

Atila Yilmaz
*Hacettepe University*
*Electrical and Electronics Engineering*
Ankara, Turkiye
ayilmaz@hacettepe.edu.tr

*Abstract*—Cardiovascular system diseases can be identified by using a specialized diagnostic process utilizing a digital stethoscope. Digital stethoscopes provide phonocardiography (PCG) recordings for further inspection, besides filtering and amplification of heart sounds. In this paper, a framework that is useful to develop feature extraction and classification of PCG recordings is presented. This framework is built upon a previously proposed segmentation algorithm that processes a feature vector produced by the agglutinate application of Mel-frequency cepstrum and discrete wavelet transform (DWT). The performance of the segmentation algorithm is also tested on a new data set and compared to the previously reported results. After identifying the fundamental heart sounds and segmenting the PCG recordings, five principal features are extracted from the time domain signal and Mel-frequency cepstral coefficients (MFCC) of each cardiac cycle. Classification outcomes are reported for three nonlinear models: *k* nearest neighbor (*k*-NN), support vector machine (SVM), and multilayer perceptrons (MLP) classifiers in comparison with a linear approach, namely Mahalanobis distance linear classifier. The results underline that although neural networks and linear classifier show compatible performance in basic classification problems, with the increase in the nonlinearity of the classification problem their performance significantly vary.

## I. INTRODUCTION

Heart sounds convey vital clues that can be evaluated through a careful auditory diagnosis. A Stethoscope is an innate part of this diagnostic method, which is also called auscultation of the heart. Although auscultation might provide easy to access information about cardiovascular system functionality, it is a highly subjective, hence limited to the various qualifications of the practitioner such as experience, hearing skills etc. [1]. Turning phonocardiography recordings into subject of digital signal processing studies may provide a prolific fundamental for future medical decision support systems that are expected to take advantage of the information concealed in heart sounds, one may not easily distinguish through conventional auscultation practices.

A cardiac cycle consists of two fundamental heart sounds and two intervals. The interval that takes place after S1 (lub) sound and before S2 (dub) sound is called systole whereas the interval after the S2 sound and before the next S1 sound is called diastole. These intervals are expected to be silent in a PCG recording of a healthy heart. On the other hand, malfunction of heart valves causes sounds that appear in systole and diastole intervals. These extra sounds besides S1 and S2 are murmurs and usually an indication of a defect in a heart valve [2].

Many studies have been reported with the objective of revealing the information heart sounds can provide. Studies on the heart sounds can be roughly split into segmentation and classification works. Segmentation is the process that identifies the fundamental heart sounds and cardiac cycle intervals on PCG recordings whereas classification focuses on labeling the PCG recordings regarding the status of the cardiovascular system health [3].

It has been a commonly accepted approach to group the segmentation studies into four categories such as reference based, envelope based, future based, probabilistic methods [4]. The first group of segmentation algorithms use subsidiary signals such as electrocardiogram (ECG). Reference based methods are usually criticized concerning the hardware and signal synchronization burden [5]. Envelope based methods simply process an envelope signal obtained from the PCG signal by various means. Fundamental heart sounds are identified on the envelope signal, usually detecting, and discriminating the peak locations of the signal. One of the most representative study of this group of segmentation algorithms has been given by C. N. Gupta *et al.* in [6]. In this study, homomorphic filtering technique was used to obtain a slow varying envelope that is used for peak detection. Application of hidden Markov models (HMM) is one of the example of the probabilistic methods that constitute a basis for many heart sounds identification and segmentation algorithms. After extracting features from four different envelope signals and

Research supported by Hacettepe University Scientific Research Projects Coordination Unit. Project Number: 19954.

labeling the heart sounds on PCG recordings with the reference of synchronously recorded ECG signal, an HMM was trained for identification purposes [7]. Feature based methods depend on the extracting and processing of the distinguishing features pertinent to heart sounds. Time varying nature of heart sounds promotes the use of time-frequency analysis techniques in this class of studies. Short-time Fourier transform (STFT) is the well-known technique that can be used to generate features for heart sounds analysis [8]. Due to its constant resolution in both time and frequency, STFT is usually replaced by discrete wavelet transform (DWT), which can provide scalable resolution and can be more adaptive to the data of concern owing to the availability of various wavelets. This makes wavelet transform advantageous for heart sounds analyses [9]. Another promising feature extraction technique for audio signals is MFCC technique. According to the source-filter acoustic model, MFCC technique provides a brief representation of source sound signal and the path it passes through [10]. Considering the location of heart valves and the direction of murmur propagation in the chest, this property of MFCC technique is employed in many studies, either for segmentation or classification algorithms [11].

Classification of PCG signals is another area of study that is subject to recent researches. Considering different phenomena behind either fundamental heart sounds or the murmurs in a PCG signal, selection of source of features plays a critical role in classification performance.

Various theories have been proposed to explain the dynamics behind the S1 and S2 sounds. The most recent and widely accepted one among these theories is hemodynamics theory and it explains the mechanism that produces heart sounds and murmurs as a combination of vibration caused by the motion of the heart valves, contraction of heart muscles, and the turbulent blood flow in the chambers [12]. These different sources of acoustic signals superpose and radiate in specific directions and tissues in the thorax. This mechanism is simply associated with the source-filter model of audio signal processing and thus MFCC technique is adopted in many studies for the feature extraction stage [13]–[15].

In this study, with the objective of the design of a medical decision support system framework a heart sounds identification and classification structure is introduced. For the first part, a segmentation algorithm that processes a feature vector obtained by application of MFCC and DWT techniques [16], [17]. The segmentation performance of the algorithm is assessed on a different data set so that the proposed segmentation algorithm is tested on two different data sets. Moreover, the new data set comprises of abnormal heart sounds with two different labels enabling evaluation of the performance on abnormal PCG recordings separately. Moreover, augmenting the feature set with MFCC features is shown to increase the performance of the classification approach [18]. Finally, besides the classification performance of three neural network linear Mahalanobis distance classifier is also included to investigate the problem complexity under a linear model and effectiveness of using nonlinear models. In Fig. 1, a flow

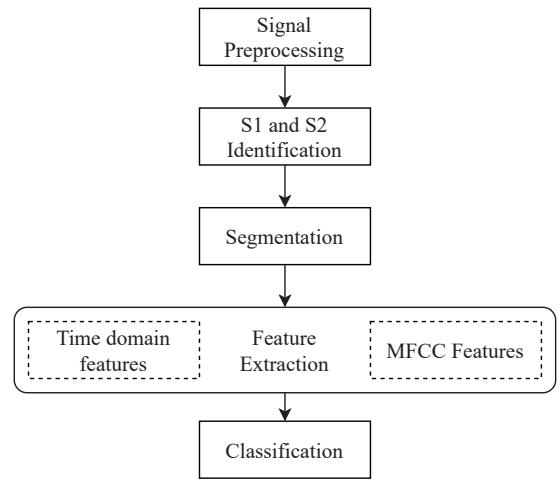

Figure 1. Signal processing work flow for segmentation and classification of PCG recordings.

diagram of the work presented in this paper is depicted briefly.

In the succeeding section, enumeration of the PCG data set is given before a brief introduction to signal preprocessing, heart sound identification and segmentation steps are provided. The segmentation performance of the algorithm is presented and compared to the results obtained in previously reported work that is accomplished on another data set. In the succeeding section, the details of the feature extraction step are introduced. Finally, classification performance utilizing linear and nonlinear classifiers will be provided. Results of the segmentation and classification works will be discussed through the perspective of a medical decision support system design in the conclusion of this paper.

## II. DATA SET

The PCG recordings are acquired in a cardiology clinic by using 3M™ Littmann® 3200 Electronic Stethoscope in diaphragm mode. A data set that consists of 150 PCG recordings is utilized to evaluate the segmentation and classification performance of the presented work in this paper. In this data set, 107 PCG recordings are labeled normal (*N*) whereas the rest are labeled abnormal (*abN*). The abnormal PCG recordings pertain to two major type of systolic heart valve diseases: 23 aortic valve stenosis (*AS*) and 20 mitral valve insufficiency (*MI*). In Table I, breakdown of these PCG recordings are presented in terms of the number of the total heart sounds and cardiac cycle.

Table I
BREAKDOWN OF THE DATA SET

| Labels | *S1* | *S2* | *Cardiac cycles* |
|---|---|---|---|
| *N* | 3232 | 3210 | 3121 |
| *AS* | 614 | 618 | 589 |
| *MI* | 574 | 571 | 552 |

## III. METHODOLOGY

To keep the organization of this paper simple and clear, only the remarkable properties of the segmentation algorithm will be mentioned in this paper. For a more detailed presentation, the readers may be referred to our previous study [16].

### A. Signal Preprocessing

In order to remove the irrelevant content in PCG recordings, the signals pass through a bandpass filter with cutoff frequencies 5 and 700 Hz. The filtered signals are normalized to their absolute maxima.

### B. S1 and S2 Identification

It is shown that replacing inverse Fourier transform with DWT in MFCC technique reduces the impact of noise and yields more robust features for S1 and S2 identification in presence of murmur [18]. This modification is applied as follows in the segmentation algorithm.

The feature vector that is obtained by application of DWT to the mel-scaled energy spectrum of the PCG signal. Warped energy spectrum of the signal is passed through Mel-filter bank. A Daubecihes - 8 (db8) type wavelet is preferred for DWT application on filter coefficients. The fifth detail coefficients (cd5) of the filter are kept as feature vector for heart sounds identification step.

Autocorrelation series of the feature vector is used for estimation of average cardiac cycle and systole durations. The feature vector is subject to a peak detection algorithm that utilizes dynamically changing thresholds. The location of the peaks, cardiac cycle and systole durations are used to construct a reference frame signal for fundamental heart sounds identification on the feature vector. Finally, an error correction step is devised to correct the erroneously labeled heart sounds. An example result of the heart sound labels, the adaptive thresholds, and the peak locations (red dots on feature vector) are shown on the PCG signal and the feature vector in Fig. 2 for a PCG recording with *MI* label.

### C. Segmentation

After locating fundamental heart sounds on the PCG signal, the algorithm segments the recording into cardiac cycle components. To limit S1 sounds a threshold value is set to the mean value of the envelope of the PCG signal. Another threshold value is set to limit S2 sounds. The later threshold is equal to the mean of the envelope of the first derivative of the signal. The intersection of the threshold and the corresponding envelopes set the borders of S1 and S2 in each cardiac and this concludes the segmentation step. In Fig. 3, an example of the segmentation result can be seen.

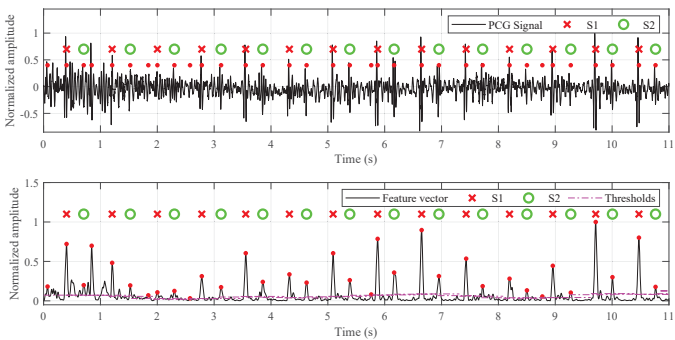

Figure 2. An example of heart sound identification on a PCG recording with mitral valve insufficiency label.

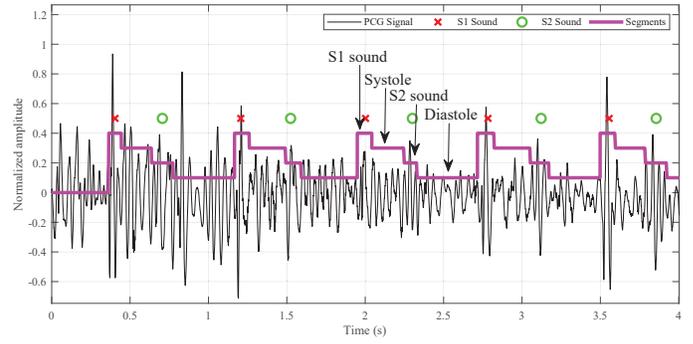

Figure 3. Result of segmentation after heart sounds identification for the same PCG recording in Fig. 2.

### D. Feature Extraction

After segmenting cardiac cycles into four regions, the first group of feature set is obtained on PCG signal. This constitutes the time domain features of the set. The second group of features are obtained from MFCC of each cardiac cycle. Below are the five principal parameters that feature set is based on: rms, variance, energy, kurtosis, dynamic interval defined as the difference between maximum and minimum amplitude of the signal of concern.

For the time domain features, the systole and the diastole intervals are further split into three subregions to reflect the difference between the shapes and the timing of different murmur types. Thus, a cardiac cycle is split into eight intervals in total for the time domain feature generation.

Extracting five principal features on each interval yields a feature vector of length 40. Different from time-domain feature, MFCC features are extracted for a full cardiac cycle. Five principal features are obtained on 12 MFC coefficients, yielding a feature vector of length 60. Combining two groups of features results in a feature vector of length 100 in total.

### E. Classification

A two-level approach depicted in Fig. 4 is proposed and found useful for classification to add more abnormalities in modular form planned in our future studies. In the first level, the PCG recordings are classified into *N* and *abN* classes. The *abN* labeled PCG recordings further classified into *AS* and *MI* classes in the second level.

Three different neural networks and linear Mahalanobis distance classifier are chosen for performance evaluation at this stage. These three neural networks are *k*-NN, SVM and MLP-BP networks. *k*-NN classifier is structured with Minkowski distance metric of third exponent whereas number of neighbors is equal to five. SVM classifier is trained with a third order polynomial kernel function. MLP-BP network has two hidden layers with nineteen and eleven neurons, respectively. Seventy percent of the feature set from each class are chosen randomly for training. Fifteen percent of the set is used for validation in MLP training whereas thirty percent of feature set is used for test in $k$-NN, SVM classifiers and linear Mahalanobis distance classifier.

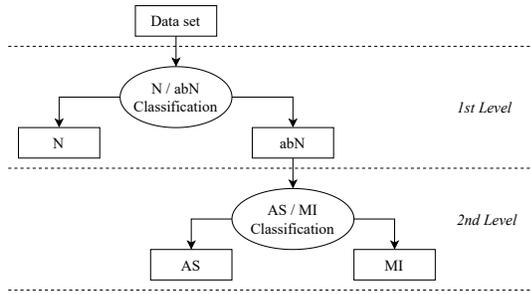

Figure 4. Two-level classification approach.

## IV. PERFORMANCE EVALUATION

In the first part of this section, the results of the segmentation and classification performance will be presented. The segmentation algorithm was previously presented in [19]. Since the segmentation algorithm is tested on a new data set in this work, its performance on the previous data set will also be provided to compare the ability of heart sounds identification and segmentation on different data sets.

In the second part, classification results will be provided comparing them with the results obtained in [19]. The feature set in this paper is augmented with cepstral-based features, thus different from [19]. Another improvement in classification work is performance of neural networks are also compared to that of the linear classifier.

### A. Segmentation Outcomes

In Table II, the performance of the segmentation algorithm is presented on the new data set. It can be concluded that the segmentation algorithm performs similarly on two different data sets comparing these outcomes with those presented in [17], where the results are obtained on an open source PCG data set available in [20]. Sensitivity is reported to be 99,51 for normal and 97,28 for abnormal recordings, whereas the precision metric is 97,59 for normal and 92,53 for abnormal PCGs [17].

Table II
EVALUATION OF THE SEGMENTATION ALGORITHM

| Performance   | Normal PCG | | Abnormal PCG | |
| criteria (%)  | S1    | S2    | S1    | S2    |
|---------------|-------|-------|-------|-------|
| Sensitivity   | 99,38 | 99,70 | 96,78 | 97,64 |
| Precision     | 99,35 | 98,99 | 96,12 | 96,53 |

Table III
FIRST AND SECOND LEVEL CLASSIFICATION PERFORMANCE OF LINEAR CLASSIFIER WITH TIME DOMAIN FEATURES

| Performance   | 1st level | | 2nd Level | |
| criteria (%)  | N     | abN   | AS    | MI    |
|---------------|-------|-------|-------|-------|
| Sensitivity   | 98,72 | 98,95 | 72,69 | 55,22 |
| Precision     | 99,62 | 96,54 | 63,02 | 65,98 |
| Specificity   | 98,95 | 98,72 | 55,22 | 72,69 |
| Accuracy      | 98,78 | | 64,16 | |

### B. Classification Outcomes

The classification performance is evaluated with the time domain features in the previous study [19]. In this study, the same approach of feature extraction is followed, and the feature set is augmented with MFCC features. Also, besides improved feature set, performance of the neural network classifiers will be compared with the performance of linear Mahalanobis distance classifier. To do so, linear Mahalanobis distance classifier is selected.

*1) Classification with time domain features:* The performance metrics for the first and the second level classification with the time domain features for linear classifier are given in Table III. Performance of the neural networks with the same feature set at the first level of classification is presented in Table IV. At the first level, both linear and nonlinear classifiers show comparable performance. Also, it is noteworthy that outcomes for three neural networks are similar to each other.

At the second level of classification, the PCG recordings are to be classified into two abnormality classes. Considering Table III and Table V together, it can be observed that each neural network classifier outperforms the linear classifier. Another remarkable change at this level is that the performance of the neural networks diverge.

*2) Classification with time domain and cepstral-based features:* The same classification work is repeated with an improved set. In the new set, time domain features are augmented with features extracted from 12 cepstral coefficients of each full cardiac cycles.

Table IV
FIRST LEVEL CLASSIFICATION PERFORMANCE OF NEURAL NETWORKS WITH TIME DOMAIN FEATURES

| Performance   | MLP   | | k-NN  | | SVM   | |
| criteria (%)  | N     | abN   | N     | abN   | N     | abN   |
|---------------|-------|-------|-------|-------|-------|-------|
| Sensitivity   | 99,97 | 99,79 | 99,89 | 100   | 100   | 99,76 |
| Precision     | 99,90 | 99,93 | 100   | 99,76 | 99,89 | 100   |
| Specificity   | 99,79 | 99,97 | 1000  | 99,89 | 99,76 | 100   |
| Accuracy      | 99,91 | | 99,93 | | 99,93 | |

Table V
SECOND LEVEL CLASSIFICATION PERFORMANCE OF NEURAL NETWORKS WITH TIME DOMAIN FEATURES

| Performance   | MLP   | | k-NN  | | SVM   | |
| criteria (%)  | AS    | MI    | AS    | MI    | AS    | MI    |
|---------------|-------|-------|-------|-------|-------|-------|
| Sensitivity   | 95,75 | 97,72 | 90,06 | 61,35 | 85,58 | 88,42 |
| Precision     | 97,55 | 96,03 | 70,97 | 85,47 | 87,52 | 86,60 |
| Specificity   | 97,72 | 95,75 | 61,35 | 90,06 | 88,42 | 88,58 |
| Accuracy      | 96,76 | | 76,05 | | 87,04 | |

Table VI
FIRST AND SECOND LEVEL CLASSIFICATION PERFORMANCE OF LINEAR CLASSIFIER WITH TIME DOMAIN AND CEPSTRAL FEATURES

| Performance criteria (%) | 1$^{st}$ level | | 2$^{nd}$ Level | |
|---|---|---|---|---|
| | N | abN | AS | MI |
| Sensitivity | 99,48 | 99,55 | 78,95 | 73,19 |
| Precision | 99,84 | 98,58 | 75,58 | 76,89 |
| Specificity | 99,55 | 98,48 | 73,19 | 78,95 |
| Accuracy | 99,50 | | 76,14 | |

The linear classifier shows improvement to some extent in systolic murmur classification after augmenting the feature vector with cepstral features. It also performs similarly with the neural networks at the first level of classification. This can be concluded by comparing Table VI with Table VII and Table VIII.

Neural networks show significant improvement especially at the second level with new feature set. Similar to the results obtained with the time domain features in Table V, performances vary discernibly at the second level of classification with the new feature set as in Table VIII.

## V. CONCLUSION

In this study, by introducing a two level-classification structure, complexity of different classification scenarios are exercised. A linear classifier and three different neural networks are employed to show that it can be advantageous to use simpler classifiers in proper classification problems. On the other hand, for some other classification problems due to increase in nonlinearity of the problems, more sophisticated classifiers can be preferred. Besides complexity of the problem and classifier choice, source of features also plays a critical role. The second level of classification is a good example of increase in the complexity since they are both systolic murmurs, thus emerges in the same interval of the cardiac cycle.

Table VII
FIRST LEVEL CLASSIFICATION PERFORMANCE OF NEURAL NETWORKS WITH TIME DOMAIN AND CEPSTRAL FEATURES

| Performance criteria (%) | MLP | | k-NN | | SVM | |
|---|---|---|---|---|---|---|
| | N | abN | N | abN | N | abN |
| Sensitivity | 100 | 100 | 100 | 100 | 100 | 100 |
| Precision | 100 | 100 | 100 | 100 | 100 | 100 |
| Specificity | 100 | 100 | 100 | 100 | 100 | 100 |
| Accuracy | 100 | | 100 | | 100 | |

Table VIII
SECOND LEVEL CLASSIFICATION PERFORMANCE OF NEURAL NETWORKS WITH TIME DOMAIN AND CEPSTRAL FEATURES

| Performance criteria (%) | MLP | | k-NN | | SVM | |
|---|---|---|---|---|---|---|
| | AS | MI | AS | MI | AS | MI |
| Sensitivity | 97,96 | 98,43 | 86,32 | 74,72 | 93,28 | 91,84 |
| Precision | 97,48 | 97,79 | 78,20 | 83,89 | 92,31 | 92,89 |
| Specificity | 98,43 | 97,96 | 74,72 | 86,32 | 91,84 | 93,28 |
| Accuracy | 98,15 | | 80,64 | | 92,57 | |

Considering problem complexity, set of features, and choice of classifiers; a multi-level and modular classification approach found to yield significant advantages for classification problems in implementation of a medical decision support system.

REFERENCES

[1] M. E. Tavel, "Cardiac auscultation - A glorious past - And it does have a future!" Circulation, vol. 113, pp. 1255-1259, Mar. 2006.
[2] Rishniw, "Murmur grading in humans and animals: past and present," Journal of Veterinary Cardiology, vol. 20, no. 4, pp. 223-233, 2018.
[3] R. Kahankova, M. Mikolasova, R. Jaros, K. Barnova, M. Ladrova and R. Martinek, "A review of recent advances and future developments in fetal phonocardiography," in IEEE Reviews in Biomed. Eng., vol. 16, pp. 653-671, 2023.
[4] M. G. Manisha Milani, Pg Emeroylariffion Abas, L. C. De Silva, "A critical review of heart sound signal segmentation algorithms," Smart Health, vol. 24, 2022.
[5] A. K. Dwivedi, S. A. Imtiaz, and E. Rodriguez-Villegas, "Algorithms for automatic analysis and classification of heart sounds - a systematic review," IEEE Access, vol. 7, pp. 8316-8345, 2018.
[6] C. N. Gupta, R. Palaniappan, S. Rajan, S. Swaminathan, and S. M. Krishnan, "Segmentation and classification of heart sounds," Canadian Conference on Elect. Comput. Eng., Saskatoon, SK, Canada, 2005.
[7] S. E. Schmidt, E. Toft, C. Holst-Hansen, C. Graff, and J. J. Struijk, "Segmentation of heart sound recordings from an electronic stethoscope by a duration dependent Hidden-Markov Model," 2008 Computers in Cardiology, Bologna, Italy, 2008.
[8] Alonso-Arvalo, Miguel A., et al. "Robust heart sound segmentation based on spectral change detection and genetic algorithms," Biomedical Signal Processing and Control, vol. 63, 2021.
[9] T. H. Chowdhury, K. N. Poudel and Y. Hu, "Time-Frequency analysis, denoising, compression, segmentation, and classification of PCG signals," in IEEE Access, vol. 8, pp. 160882-160890, 2020.
[10] S. Paulose, D. Mathew, A. Thomas, "Performance evaluation of different modeling methods and classifiers with MFCC and IHC features for speaker recognition," Procedia Computer Science, vol. 115, 2017.
[11] N. Shreyas, M. Venkatraman, S. Malini, and S. Chandrakala, "Trends of sound event recognition in audio surveillance: A recent review and study," The Cognitive Approach in Cloud Computing and Internet of Things Technologies for Surveillance Tracking Systems, 2020.
[12] S. Crandon, M. S. Elbaz, J. J. Westenberg, R. J. van der Geest, S. Plein, and P. Garg, "Clinical applications of intracardiac four-dimensional flow cardiovascular magnetic resonance: A systematic review," International journal of cardiology, vol. 249, pp. 486-493, 2017.
[13] T. Fernando, H. Ghaemmaghami, S. Denman, S. Sridharan, N. Hussain and C. Fookes, "Heart sound segmentation using bidirectional LSTMs with attention," in IEEE J. of Biomed. and Health Informat., vol. 24, no. 6, pp. 1601-1609, June 2020.
[14] S. Das, S. Pal, M. Mitra, "Acoustic feature based unsupervised approach of heart sound event detection," Comput. in Biology and Med., vol. 126, 2020.
[15] S. Prusty, S. Patnaik and S. K. Dash, "Differentiating S1, S2 noises from abnormal heart sounds generated in closure of atrioventricular and semilunar Valves using MFCC and LSTM," 2022 1st IEEE International Conference on Industrial Electronics: Developments and Applications (ICIDeA), Bhubaneswar, India, 2022, pp. 208-213.
[16] I. Ozkan, A. Yilmaz, and G. Celebi, "Hybrid segmentation algorithm using mel-frequency cepstrum and wavelet transform for phonocardiography records," 27th Signal Processing and Communications Applications Conference (SIU), Sivas, Turkey, 2019.
[17] I. Ozkan, A. Yilmaz, and G. Celebi, "Improved segmentation with dynamic threshold adjustment for phonocardiography eecordings," 41st Annual International Conference of the IEEE Engineering in Medicine and Biology Society (EMBC), Berlin, Germany, 2019, pp. 6681-6684.
[18] K. Courtemanche, V. Millette, and N. Baddour, "Heart sound segmentation based on mel-scaled wavelet transform," The Canadian Medical and Biological Engineering Society Proceedings, vol. 31, no. 1, 2008.
[19] I. Ozkan and A. Yilmaz, "Classification of systolic murmurs by using improved mel-wavelet segmentation," 29th Signal Processing and Communications Applications Conference (SIU), Istanbul, Turkey, 2021.
[20] P. Bentley, G. Nordehn, M. Coimbra, S. Mannor, "The PASCAL classifying heart sounds challenge 2011 Results," [Online]. Available: http://www.peterjbentley.com/heartchallenge/index.htm